\RequirePackage[l2tabu, orthodox]{nag}

\documentclass[conference]{IEEEtran}

\usepackage[T1]{fontenc}              % 8-bit font encoding
\usepackage[utf8]{inputenc}           % Should be default, but let's be safe
\usepackage[nopatch=eqnum]{microtype} % Subliminal refinements towards typographical perfection
\usepackage[scaled=.8]{beramono}      % Nice-looking monospace font
\usepackage[abbreviations]{foreign}   % \eg, \ie, etc.
\usepackage[all]{nowidow}
\usepackage{listings}
\usepackage{hyperref}
\usepackage{graphicx}
\usepackage{booktabs}
\usepackage{xcolor}
\usepackage{siunitx}
\usepackage{subcaption}
\usepackage{pifont}
\usepackage{amsmath}
\usepackage{amssymb}
\usepackage{balance}
\usepackage[multiple]{footmisc}
\usepackage[most]{tcolorbox}
\usepackage[numbers]{natbib}

\usepackage{cleveref}                 % The almighty \Cref command, keep in last position

\hypersetup{
	breaklinks,
    colorlinks,
    allcolors=black
}

\newcommand{\rs}{\textsc{Roseau}\xspace}
\newcommand{\rss}{\textsc{RoseauJDT}\xspace}
\newcommand{\rsb}{\textsc{RoseauASM}\xspace}

\newcommand{\japi}{\textsc{Japicmp}\xspace}
\newcommand{\rapi}{\textsc{Revapi}\xspace}

\newcommand{\fs}{F\textsubscript{1}-score\xspace}

\definecolor{keywordscolor}{RGB}{0, 51, 179}
\definecolor{stringcolor}{RGB}{6, 125, 23}
\definecolor{commentscolor}{RGB}{140, 140, 140}
\definecolor{annotationscolor}{RGB}{100, 100, 100}
\definecolor{lstbgcolor}{RGB}{251, 251, 251}

\newcommand*{\ijava}[1]{\lstinline[basicstyle=\ttfamily,language=Java,keywordstyle=\mdseries\color{keywordscolor},morekeywords={sealed,record}]{#1}}
\newcommand*{\etclst}{\colorbox{blue!10}{\textbf{\vphantom{a}\dots}}}

\newtcolorbox{finding}{
	colback=gray!10,    % Light gray background
	colframe=black,     % Black border
	boxrule=0.8pt,      % Border thickness
	arc=3pt,            % Rounded corners
	left=6pt,           % Padding on the left
	right=6pt,          % Padding on the right
	top=6pt,            % Padding on the top
	bottom=6pt,         % Padding on the bottom
	enhanced            % Enables advanced styling
}

\begin{document}

\title{\rs: Fast, Accurate, Source-based\\API Breaking Change Analysis in Java}

\author{
    \IEEEauthorblockN{
        Corentin Latappy,\IEEEauthorrefmark{1}
        Thomas Degueule,\IEEEauthorrefmark{1}
        Jean-Rémy Falleri,\IEEEauthorrefmark{1}\IEEEauthorrefmark{3}
        Romain Robbes\IEEEauthorrefmark{1} and
        Lina Ochoa\IEEEauthorrefmark{2}
    }
    \IEEEauthorblockA{\IEEEauthorrefmark{1}
        Univ. Bordeaux, CNRS, Bordeaux INP, LaBRI, UMR 5800,
        Talence, France\\
        Email: firstname.lastname@labri.fr
    }
    \IEEEauthorblockA{\IEEEauthorrefmark{2}
        Eindhoven University of Technology,
        Eindhoven, The Netherlands\\
        Email: l.m.ochoa.venegas@tue.nl
    }
    \IEEEauthorblockA{\IEEEauthorrefmark{3}
        Institut Universitaire de France
    }
}

\maketitle

\begin{abstract}
Understanding API evolution and the introduction of breaking changes (BCs) in software libraries is essential for library maintainers to manage backward compatibility and for researchers to conduct empirical studies on software library evolution.
In Java, tools such as \japi and \rapi are commonly used to detect BCs between library releases, but their reliance on binary JARs limits their applicability.
This restriction hinders large-scale longitudinal studies of API evolution and fine-grained analyses such as commit-level BC detection.

In this paper, we introduce \rs, a novel static analysis tool that constructs technology-agnostic API models from library code equipped with rich semantic analyses.
API models can be analyzed to study API evolution and compared to identify BCs between any two versions of a library (releases, commits, branches, \etc).
Unlike traditional approaches, \rs can build API models from source code or bytecode, and is optimized for large-scale longitudinal analyses of library histories.

We assess the accuracy, performance, and suitability of \rs for longitudinal studies of API evolution, using \japi and \rapi as baselines.
We extend and refine an established benchmark of BCs and show that \rs achieves higher accuracy ($F_1 = 0.99$) than \japi ($F_1 = 0.86$) and \rapi ($F_1 = 0.91$).
We analyze 60 popular libraries from Maven Central and find that \rs delivers excellent performance, detecting BCs between versions in under two seconds, including in libraries with hundreds of thousands of lines of code.
We further illustrate the limitations of \japi and \rapi for longitudinal studies and the novel analysis capabilities offered by \rs by tracking the evolution of Google's Guava API and the introduction of BCs over 14 years and 6,839 commits, reducing analysis times from a few days to a few minutes.
\end{abstract}


\section{Introduction}
\label{sec:intro}

Like any other software, software libraries continuously evolve to introduce new features, fix bugs, enhance security and performance, and implement refactorings.
Library users must keep their dependencies up to date to take advantage of these improvements and prevent technical debt build-up~\cite{zerouali2019formal}.
However, when a library evolves, it may introduce breaking changes (BCs) that disrupt backward compatibility with existing clients.
Syntactic BCs, in particular, affect the Application Programming Interface (API) of libraries, manifesting as compilation and linking errors in client code.
Consequently, clients may delay upgrading their dependencies, raising security concerns and complicating future updates~\cite{kula2018developers,mirhosseini2017can}.

Automatic BC detection tools have been proposed to help library developers identify and document these changes~\cite{jezek2017api}.
In particular, tools such as \japi and \rapi are actively used in the development of popular Java libraries and frameworks such as Apache Commons~\cite{japicmp_report_apache}, Spring Framework~\cite{japicmp_report_spring}, Google Gson~\cite{japicmp_config_gson}, and Neo4j~\cite{japicmp_config_neo4j}. These tools also enable research contributions and empirical studies of software evolution.
Notably, they have been used to identify BCs in large software repositories~\cite{ochoa2022breaking,jayasuriya2025extended}, analyze breaking dependency updates~\cite{reyes2024breaking}, construct safe migration plans for outdated dependencies~\cite{dann2023upcy,jaime2024balancing}, analyze software versioning schemes~\cite{zhang2022has}, and assist code reviews in pull requests~\cite{ochoa22breakbot}.

However, existing BC detection tools can only consume binary JAR archives and cannot directly analyze the source code of software libraries.
Using them requires compiling the library versions of interest, which significantly hinders their use in several key scenarios.
For practitioners, this implies that backward compatibility checks must be delayed until after a binary is built, which can be unnecessarily costly in continuous integration.
This also prevents their use for immediate feedback in IDEs and for quickly identifying commits or pull requests that introduce backward-incompatible changes~\cite{ochoa22breakbot}.
For researchers, empirical evidence suggests that nearly half of Java software projects hosted on GitHub cannot be built automatically, restricting the libraries they can analyze with these tools~\cite{hassan2017automatic,sulir2020large,jayasuriya2025extended}.
Even when projects can be successfully built, compilation is costly and impractical for large-scale longitudinal studies of API evolution.
Consequently, researchers have primarily focused on analyzing library releases readily available in package managers, overlooking research questions at the finer granularity of individual changes and commits.

In this paper, we present a new approach to API analysis and syntactic BC detection that reifies API models as first-class entities.
API models are designed to be self-contained and technology-independent, allowing extraction from any suitable source, such as source code, bytecode, or even IDE-specific internal code representations~\cite{jetbrains_psi}.
BCs between two library versions can be inferred by comparing their respective API models using straightforward detection rules.
We implement our approach in a new tool, \rs, which emphasizes accuracy and performance to enable large-scale and fine-grained longitudinal studies of API evolution.
We evaluate the accuracy, performance, and suitability of \rs for longitudinal studies using \japi and \rapi as baselines.

% We evaluate \rs with respect to \japi and \rapi using the following research questions:
% \begin{description}
%     \item[RQ1] \textit{How accurately does \rs identify BCs in Java library APIs?}\\
%     We evaluate \rs’s precision and recall compared to \japi and \rapi on an established BCs benchmark~\cite{jezek2017api}.
%     \item[RQ2] \textit{How quickly does \rs infer BCs?}\\
%     We measure the performance of \rs on a diverse dataset of popular Java libraries compared to \japi and \rapi.
%     \item[RQ3] \textit{Can \rs be effectively applied to analyze library histories?}
%     We investigate whether \rs enables longitudinal analyses of library histories.
% \end{description}
Specifically, this paper makes the following contributions:
\begin{itemize}
    \item We present a novel approach for inferring accurate API models from source code or bytecode, and for detecting BCs by comparing API models;
    \item We present an implementation of this approach in \rs, a new BC detection tool optimized for accuracy and performance;
    \item We extend and refine an established benchmark of BCs proposed by \citeauthor{jezek2017api}~\cite{jezek2017api} with new metrics, stronger oracles, and a streamlined evaluation pipeline;
    \item We evaluate \rs, \japi, and \rapi on the extended benchmark, and find that \rs achieves higher precision and recall ($F_1 = 0.99$) than \japi ($F_1 = 0.86$) and \rapi ($F_1 = 0.91$);
    \item We evaluate \rs's runtime performance and find that it matches \japi and surpasses \rapi when analyzing binary JARs. When analyzing source code and commit histories, \rs outperforms both by two orders of magnitude, as it does not require compiling source code;
    \item We showcase \rs's suitability for large-scale longitudinal studies of API evolution by analyzing 14 years and 6,839 commits of the popular Guava library, reducing analysis times from a few days to a few minutes.
\end{itemize}

The remainder of this paper is organized as follows.
We review related work and identify the limitations of current BC detection tools in \Cref{sec:background}.
We present API models and BC detection rules in \Cref{sec:approach}, and their optimized implementation in \rs in \Cref{sec:roseau}.
We evaluate \rs's accuracy, performance, and suitability for large-scale longitudinal studies in \Cref{sec:evaluation}, using \japi and \rapi as baselines.
Finally, we conclude the paper in \Cref{sec:conclusion}.

\section{Related Work and Motivation}
\label{sec:background}

Modern software system development typically involves the use of third-party software libraries. 
Application Programming Interfaces (APIs) govern the interactions between a library and its clients, constituting the primary friction point during evolution.
Library developers must carefully design and evolve their APIs so that clients can benefit from new features, bug fixes, or performance improvements without having to constantly rewrite their code.
In particular, introducing BCs in an API burdens client developers, who may hesitate to upgrade their dependencies, raising security concerns as they might miss security patches, and making future upgrades even more difficult as client updates accumulate.

In this section, we review some background notions on backward compatibility and BCs and discuss the importance of BCs in both software engineering practice and research.

\subsection{Backward Compatibility and Breaking Changes}

Whenever a library evolves, it may break the contract previously established with its clients by introducing BCs.
BCs are language-specific and can either be \emph{syntactic} or \emph{semantic} in nature~\cite{jayasuriya2025extended}.
Syntactic changes affect the library's API, \ie the signatures of exported symbols accessible from client code (types, methods, etc.).
Semantic (or behavioral) changes affect the run-time behavior of a library and can alter client functionalities or trigger run-time errors~\cite{jayasuriya2024understanding}.
Changes that affect neither the API nor the behavior of software libraries are backward-compatible.

In Java, syntactic BCs are further categorized into source-incompatible and binary-incompatible changes.
Source-incompatible changes occur when recompiling an existing client against a new version of a library results in compilation errors.
Binary-incompatible changes occur when linking a pre-existing client binary with a new library binary results in class loading or linking errors~\cite[\S13.2]{jls}.
Some BCs are only source-incompatible (\eg a method throwing a new type of checked exception), some others are only binary-incompatible (\eg changing the return type of a method covariantly), and some are both source- and binary-incompatible (\eg removing a method)~\cite{kinds_of_compatibility}.

\begin{figure}[bt]
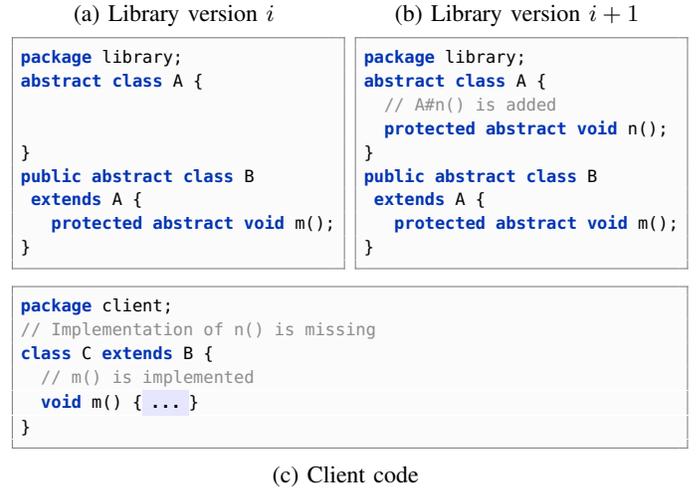

\centering
\begin{subfigure}[bt]{.485\linewidth}
\centering
\caption{Library version $i$}
\label{lst:v1}
\begin{lstlisting}[language=Java]
package library;
abstract class A {


}
public abstract class B 
 extends A {
   protected abstract void m();
}

\end{lstlisting}
\end{subfigure}
\hfill
\begin{subfigure}[bt]{.485\linewidth}
\centering
\caption{Library version $i+1$}
\label{lst:v2}
\begin{lstlisting}[language=Java]
package library;
abstract class A {
  // A#n() is added
  protected abstract void n();
}
public abstract class B 
 extends A { 
   protected abstract void m();
}
\end{lstlisting}
\end{subfigure}
\begin{subfigure}[bt]{\linewidth}
\centering
\begin{lstlisting}[language=Java,belowskip=0pt]
package client;
// Implementation of n() is missing
class C extends B {
  // m() is implemented
  void m() {!*\etclst*!}
}
\end{lstlisting}
\caption{Client code}
\end{subfigure}
\caption{An example BC propagating to clients.}
\label{fig:bc}
\end{figure}

Identifying changes that affect APIs is not always straightforward.
While some BCs are overt (such as removing a public type from an API), others are more subtle.
\Cref{fig:bc} depicts an example of the latter, highlighting the importance of accurately delimiting the scope of an API.
In this example, the abstract class \ijava{B} extends an internal class \ijava{A}.
An external client extends \ijava{B} via the concrete class \ijava{C}.
The library maintainers add an abstract method to \ijava{A} that leaks into \ijava{C}, raising a missing implementation error, although there is no direct use of the modified internal class \ijava{A}.
% the method \ijava{m()} defined in class \ijava{A}, which is not part of the library's API, is leaked to clients through the type \ijava{B} that re-exports it to subtypes.
% Class \ijava{C} in client code can access \ijava{m()} through subclassing and is thus vulnerable to the BC introduced in version 2.
In this case, maintainers may implement this change without realizing its potential impact on clients.
Conversely, maintainers may hesitate to make changes even when they would not break backward compatibility.
To alleviate these problems, several tools have been proposed to assist maintainers in making informed decisions when evolving their APIs.
%, which we introduce in the next section.

%\begin{itemize}
%    \item Backward compatibility is i) important ii) hard to achieve
%    \item Some BCs are obvious, others aren't; accurate delimitation of APIs is crucial
%    \item Languages evolve (recently: sealed, records, \etc) and so does the definition of BCs
%    \item Various interpretations, importance of communication and policies~\cite{bogart2021and}
%\end{itemize}

% \subsection{Breaking Change Detection Tools}
% \label{subsec:breaking_change_detection_tools}

\subsection{Breaking Changes in Software Engineering Practice}

Although most ecosystems and library maintainers aim for stability and compatibility throughout evolution, their policies regarding BCs management can vary~\cite{bogart2021and}.
For instance, maintainers of the popular Apache Commons libraries prohibit breaking binary compatibility in patch and minor releases.
In case of binary-incompatible major releases, they recommend changing the package name and Maven coordinates of the library~\cite{apache_commons_binary_compa}.
Google's Guava attempts to maintain backward compatibility across releases, except for APIs explicitly marked as unstable~\cite{guava_github}.
Some libraries also use various methods to mark specific APIs as unstable, such as using \texttt{@Beta} and \texttt{@Experimental} annotations or placing these APIs in \texttt{*.internal.*} packages.

To implement these policies, different ecosystems offer (breaking) change detection tools, for instance, in Rust (\texttt{cargo-semver-checks}), Go (\texttt{gocompat} and \texttt{gorelease}), and Python (AexPy~\cite{du2022aexpy}).
The Java ecosystem, in particular, boasts a rich set of BC detection tools, inventoried and evaluated by \citeauthor{jezek2017api}~\cite{jezek2017api}.
Although many have been abandoned over time, \japi\footnote{https://siom79.github.io/japicmp/} and \rapi\footnote{https://revapi.org/revapi-site/main/} are still actively maintained.
\japi and \rapi take as input two versions of a library as binary JARs and diff them to identify both source-incompatible and binary-incompatible changes.
% The Java Language Specification plays an important role in how these changes are identified, as it provides a whole chapter on binary compatibility~\cite[\S13]{jls}.
They produce reports in various formats and can be integrated into continuous integration through dedicated build manager plug-ins.
These tools are used in the development of numerous libraries such as Apache's \texttt{commons-lang}~\cite{japicmp_report_apache}, Spring's \texttt{spring-framework}~\cite{japicmp_report_spring}, and Google's \texttt{gson}~\cite{japicmp_config_gson}.
Other libraries, such as Apache's Hadoop and Google's Guava use simpler API differencing tools, such as JDiff, to list API differences without distinguishing backward-compatible changes and BCs~\cite{japicmp_report_guava}.

%Several tools have been implemented in different ecosystems to assist developers in identifying BCs and inform their decisions regarding API evolution.
% Different ecosystems offer various tools to assist library maintainers, for instance, in Rust (\texttt{cargo-semver-checks}), Go (\texttt{gocompat} and \texttt{gorelease}), and Python (AexPy~\cite{du2022aexpy}) to support BC detection.
% In Rust, the \texttt{cargo-semver-checks} tool checks for BCs in new versions of Rust packages to determine whether they align with semantic versioning conventions. %\footnote{\url{https://github.com/obi1kenobi/cargo-semver-checks}}
% In Go, tools such as \texttt{gocompat} and \texttt{gorelease} analyze changes introduced in public Go APIs.
% Tools such as AexPy~\cite{du2022aexpy} can be used to find BCs in Python packages.

%In practice, existing tools have several limitations due to their reliance on binary JARs.
The aforementioned tools are well-suited for documenting BCs between releases.
However, their reliance on binary JARs introduces several limitations.
They incur a substantial cost in continuous integration, as they require compiling the library and delaying BC detection until after binary construction.
These steps could be avoided if backward compatibility could be verified at the source code level.
Furthermore, they typically use a published version of the library as a reference, which is compared against the current development version.
Therefore, branch comparisons or pull request reviews become arduous, as the reference development version is usually unreleased and thus not easily referenced.
This highlights the need for an efficient source-level approach to detecting BCs.

\subsection{Breaking Changes in Software Engineering Research}

API evolution is a thriving research area that has garnered considerable interest in the literature, particularly through empirical studies on software ecosystems~\cite{lamothe2021systematic}.
BCs are central to API evolution, and several studies have investigated their prevalence, causes, and impacts, particularly in Java ecosystems (\eg, \cite{xavier2017historical,jayasuriya2023understanding,ochoa2022breaking,jayasuriya2025extended,reyes2024breaking}), with more recent research extending to npm~\cite{venturini2023depended} and the Golang ecosystem~\cite{li2023large}.
Other studies have focused on facilitating client updates while prioritizing security, minimizing technical lag, and preserving backward compatibility~\cite{dann2023upcy,zhang2023compatible}.
In the same line, \citeauthor{jaime2024balancing} model a client update as a multi-objective optimization problem accounting for potential BCs~\cite{jaime2024balancing}.

Overall, the above-mentioned studies identify BCs, often relying on \japi, and analyze their introduction and impact at the granularity of library releases.
% , for instance, to check adherence to semantic versioning principles.
\citeauthor{brito2020you}, however, study API evolution at the finer granularity of individual commits using the APIDiff tool~\cite{brito2020you}.
APIDiff can detect common refactorings in APIs (such as method pull-up) and certain BCs, but the list of supported BCs is limited compared to state-of-the-art tools such as \japi or \rapi~\cite{brito2018apidiff}.
\citeauthor{ochoa22breakbot} identified the need to analyze the introduction of BCs in individual pull requests instead of waiting for releases~\cite{ochoa22breakbot}.
However, their tool, Maracas, still relies on \japi and comes at the cost and sometimes impossibility of compiling the two branches.

The reliance of current tools on binary JARs is unsurprising, as Java bytecode is significantly more stable and less sensitive to changes in the Java language than source code, while also being straightforward and efficient to parse.
However, JARs are not always readily available, which severely hinders empirical research on the evolution of Java libraries.
Various authors report that only around half of Java projects on GitHub can be built automatically~\cite{hassan2017automatic,sulir2020large,jayasuriya2025extended}, which prevents the analysis of individual commits in the wild.
Even when the project successfully compiles, the compilation phase is too costly and impractical for large-scale studies.
Thus, researchers have focused on analyzing releases of popular libraries that are already readily available, eluding research questions that would require analyzing individual commits, branches, or pull requests.
This highlights the need for an approach that enables researchers to analyze source code directly and scales BC analysis on large datasets of commit histories.

\section{Detecting Breaking Changes with API Models}
\label{sec:approach}

Syntactic BCs are changes that affect the APIs of software libraries---contracts established between these libraries and their clients.
While modifications to internal implementations can introduce semantic BCs, they do not affect APIs.
Similarly, changes in APIs explicitly documented as internal or beta-access are outside the scope of BC detection as they are usually exempt from compatibility guarantees.
Therefore, an essential first step is to precisely delimit the scope of a library's API.
Identifying BCs between two versions then involves comparing the two APIs, cataloging their differences, and determining whether these differences compromise backward compatibility at the source or binary levels.

In our approach, we reify API models as self-contained first-class entities that can be reasoned about to track API evolution (\Cref{sec:api}).
API models can then be compared to infer a list of BCs using straightforward detection rules expressed directly at the API level (\Cref{sec:detection}).
Throughout this section, we reference the specification of the latest long-term support version of Java (JLS~21~\cite{jls}) when appropriate.

\subsection{API Models}
\label{sec:api}

Similar to prior work on syntactic usage models~\cite{monce_lightweight_2024}, we consider that a software library declares a set of \emph{symbols}.
Each symbol has a unique identifier within the library and an associated declaration that specifies its properties:~kind, visibility, modifiers, location, \etc.
The JLS~21 defines the following symbol kinds relevant to API definitions:~type declarations (classes, interfaces, enumerations, records, and annotation interfaces), executables (constructors and methods), and fields~\cite[\S6.1]{jls}.
For instance, Java's standard library declares the symbol \ijava{public class String} of kind \textit{class} and visibility \ijava{public}, located in the file \texttt{String.java}.
It also declares a field \ijava{private int hash} of type \ijava{int} and visibility \ijava{private}, and a method \ijava{public char charAt(int index)} of visibility \ijava{public} that takes an \ijava{int} as an argument and returns a \ijava{char} value.

Type declarations and fields are uniquely identified by their fully qualified name such as \texttt{java.lang.String} and \texttt{java.lang.String.hash}.
Methods are uniquely identified by the fully qualified erasure of their signatures such as \texttt{java.lang.String.charAt(int)}~\cite[\S8.4.2]{jls}.

Symbols can be \emph{exported}, allowing client code outside the library to access them.
Given a function $exported: \mathcal{S} \rightarrow \{\top, \bot\}$ that indicates whether a given symbol $s \in \mathcal{S}$ is exported, the API of a library is the set $\mathcal{A} \subseteq \mathcal{S}$ such that $\mathcal{A} = \{s \in \mathcal{S}: \mathit{exported}(s) = \top\}$.
In Java, a symbol is considered exported (or accessible) if every symbol in the visibility chain---from the compilation unit to the symbol itself---is exported~\cite[\S6.6.1]{jls}.\footnote{Recent versions of Java offer a \emph{module} construct that enables finer control over exports but is beyond the scope of this paper.}
More specifically, the following rules apply:

\begin{itemize}
    \item \textit{Public symbols:} (transitively-)\ijava{public} type declarations, as well as \ijava{public} methods and fields within \ijava{public} type declarations, can be accessed from anywhere without restriction;
    \item \textit{Protected symbols:} \ijava{protected} fields, methods, and type declarations within effectively extensible \ijava{public} type declarations can be accessed through subtyping or by code located in the same package. An effectively extensible type declaration is one that is neither \ijava{final} nor \ijava{sealed} and that, in the case of classes, possesses a \ijava{public} or \ijava{protected} constructor that subclasses can access;
    \item \textit{Package-private symbols:} package-private symbols (Java's default visibility in the absence of a visibility modifier) can be accessed by code located in the same package.
\end{itemize}

Note that the accessibility of an inherited member is assessed in the context of the inheriting type, not the original declaring type.
This is, for instance, the case in \Cref{fig:bc} where the method \texttt{n()} is ``re-exported'' to clients through the subclass \ijava{B}.

The only legal visibilities for top-level type declarations are \ijava{public} and package-private, so library developers must use the latter to prevent clients from accessing them.
In our model, we consider all \ijava{public} and \ijava{protected} symbols as exported.
Package-private symbols are not considered exported because client code intentionally using a package with the same name as the library's to access its package-private symbols is a pathological case that most likely breaches the API's intent.

% \begin{figure}[tb]
%     %\missingfigure{}
%     \includegraphics[width=\linewidth]{figures/test.drawio.pdf}
%     \caption{The API Model}
%     \label{fig:api}
% \end{figure}

Symbols are organized in a hierarchy with kind-specific properties.
Each symbol is associated with its unique identifier, visibility, modifiers, and annotations.
Different symbol kinds accept different legal modifiers.
Classes can, for example, be declared \ijava{abstract} or \ijava{sealed}, while methods can be declared \ijava{default} or \ijava{native}.
Type declarations hold fields and methods and can contain other type declarations (\emph{nested} or \emph{inner}).
They can also declare a list of type variables (aka. formal type parameters) and implemented interfaces.
In addition, classes can declare a unique superclass and a set of constructors.
Executables hold a list of parameters, a return type, a list of type variables, and potentially thrown exception types.
Fields have a specific type.
References to type declarations (\eg, a thrown exception, a super-class, a return type) are modeled as type references that associate names to declarations.

An API model is thus a graph of API symbols with a spanning containment tree representing the natural containment relationships between type declarations, fields, and executables, and cross-references linking type references to type declarations.
Note that the information contained in an API model is purely syntactic and can be trivially inferred from source code (\eg, by visiting an abstract syntax tree) or bytecode (\eg, by parsing class files).
However, raw information extracted directly from source code or bytecode is insufficient for accurate analysis.
Some properties of symbols are not always explicitly set in source code or bytecode (\eg, implicit constructors) while others must be derived (\eg, all supertypes of a type declaration).
For instance, interfaces nested within other types are implicitly static, even if the \ijava{static} modifier is not explicitly present.
A class is considered \emph{effectively final} (respectively \emph{effectively abstract}) if (i)~it is explicitly declared as \ijava{final} or \ijava{sealed} (respectively \ijava{abstract}), or (ii)~none of its constructors are accessible to subclasses.
Inferring the precise methods or fields that can be invoked or accessed on a type declaration involves resolving all supertypes in its hierarchy while accounting for Java's overriding, shadowing, and hiding semantics~\cite[\S8.4.8]{jls}.

API models automatically infer this missing and derived information that is not readily available in source code or bytecode to ensure accurate analysis, accounting for differences between source code and bytecode, as compilers typically infer certain implicit properties during compilation.
The underlying tool responsible for visiting the abstract syntax trees or class files can thus focus solely on syntax and only extract the information explicitly set in the source.

\begin{table*}[ht]
	\scriptsize
	\renewcommand{\arraystretch}{0}
	\centering
	\caption{BC detection rules implemented in \rs. $s \sim s'$ denotes that symbols $s$ and $s'$ have matching fully qualified names or erasures. \ding{51}/\ding{55} denotes whether a change is source- or binary-breaking (\ding{55}) and source- or binary-compatible (\ding{51}). Executable-related BCs apply to both methods and constructors. Annotation-related BCs are omitted for the sake of conciseness.}
	\label{tab:bcs}
	\begin{tabular}{@{}rlcc@{}}
		\textbf{Kind} & \textbf{Detection Rule} & \textbf{Binary} & \textbf{Source} \\ \midrule
		Type removed & \(T \in A \land \nexists\, T' \in A' \mid T \sim T'\) & \ding{55} & \ding{55} \\ %\midrule
		Type now protected & \(T \sim T' \land \text{public}(T) \land \text{protected}(T')\) & \ding{55} & \ding{55} \\ %\midrule
		Type kind changed & \(T \sim T' \land \text{typeKind}(T) \neq \text{typeKind}(T')\) & \ding{55} & \ding{55} \\ %\midrule
		Supertype removed & \(T \sim T' \land \exists\, S \in \text{superTypes}(T) \mid \text{exported}(S) \land T' \not\sqsubseteq S\) & \ding{55} & \ding{55} \\ %\midrule
		Type variable removed & \(T \sim T' \land |\text{typeVars}(T)| > |\text{typeVars}(T')|\) & \ding{51} & \ding{55} \\ %\midrule
		Type variable added & \(T \sim T' \land |\text{typeVars}(T')| > |\text{typeVars}(T)| > 0\) & \ding{51} & \ding{55} \\ %\midrule
		Type variable changed & 
		\(T \sim T' \land \exists i \in [1,|\text{typeVars}(T)|], \exists b' \in \text{bounds}(\text{typeVar}_i(T')), \forall b \in \text{bounds}(\text{typeVar}_i(T)) \mid b \not\sqsubseteq b'\) & \ding{51} & \ding{55} \\ %\midrule
		Type now final & \(T \sim T' \land \neg\text{effectivelyFinal}(T) \land \text{effectivelyFinal}(T')\) & \ding{55} & \ding{55} \\ %\midrule
		Type now abstract & \(T \sim T' \land \neg\text{effectivelyFinal}(T) \land \text{effectivelyFinal}(T')\) & \ding{55} & \ding{55} \\ %\midrule
		Nested type now static & 
		\(T \sim T' \land \neg\text{static}(T) \land \text{static}(T')\) & \ding{55} & \ding{55} \\ %\midrule
		Nested type no longer static & 
		\(T \sim T' \land \text{static}(T) \land \neg\text{static}(T')\) & \ding{55} & \ding{55} \\ %\midrule
		Class now checked exception & \(T \sim T' \land \text{uncheckedException}(T) \land \text{checkedException}(T')\) & \ding{51} & \ding{55} \\
        Abstract method added to type & \( T \sim T' \land \exists m' \in \text{allMethods}(T') \mid \text{abstract}(m') \land \nexists m \in \text{allMethods}(T) \mid m \sim m' \) & \ding{51} & \ding{55} \\ \midrule % \midrule
		Executable removed & 
		\(e \in \text{executables}(T) \land \nexists\, e' \in \text{executables}(T') \mid e \sim e'\) & \ding{55} & \ding{55} \\ %\midrule
		Executable now protected & 
		\(e \sim e' \land \text{public}(e) \land \text{protected}(e')\) & \ding{55} & \ding{55} \\ %\midrule
		Executable checked exception removed & 
		\(e \sim e' \land \exists\, t \in \text{thrown}(e) \mid \nexists t' \in \text{thrown}(e'),\, t' \sqsubseteq t\) & \ding{51} & \ding{55} \\ %\midrule
		Executable checked exception added & 
		\(e \sim e' \land \exists\, t' \in \text{thrown}(e') \mid \nexists\, t \in \text{thrown}(e),\, t' \sqsubseteq t\) & \ding{51} & \ding{55} \\ %\midrule
		Executable parameter generics changed & \(e \sim e' \land \exists\, i \in [1,|\text{params}(e)|] \mid |\text{typeArgs}(\text{param}_i(e))| \neq |\text{typeArgs}(\text{param}_i(e'))| \) & \ding{51} & \ding{55} \\ %\midrule
		Executable type variable removed & 
		\(e \sim e' \land |\text{typeVars}(e)| > |\text{typeVars}(e')| \land (\text{isMethod}(e) \lor |\text{typeVars}(e)| > 1)\) & \ding{51} & \ding{55} \\ %\midrule
		Executable type variable added & 
		\(e \sim e' \land |\text{typeVars}(e')| > |\text{typeVars}(e)| > 0 \) & \ding{51} & \ding{55} \\ %\midrule
		Executable type variable changed & \(e \sim e' \land \exists\, i \in [1,|\text{typeVars}(e)|], \exists\, b' \in \text{bounds}(\text{typeVar}_i(e')), \forall\, b \in \text{bounds}(\text{typeVar}_i(e)) \mid b \not\sqsubseteq b'\) & \ding{51} & \ding{55} \\ \midrule
        Method now final & 
		\(m \sim m' \land \neg\text{effectivelyFinal}(m) \land \text{effectivelyFinal}(m')\) & \ding{55} & \ding{55} \\ %\midrule
		Method now static & 
		\(m \sim m' \land \neg\text{static}(m) \land \text{static}(m')\) & \ding{55} & \ding{51} \\ %\midrule
		Method no longer static & 
		\(m \sim m' \land \text{static}(m) \land \neg\text{static}(m')\) & \ding{55} & \ding{55} \\ %\midrule
		Method now abstract & 
		\(m \sim m' \land \neg\text{abstract}(m) \land \text{abstract}(m')\) & \ding{55} & \ding{55} \\ %\midrule
		Method return type changed & 
		\(m \sim m' \land \text{returnType}(m) \neq \text{returnType}(m')\) & \ding{55} & \ding{55} \\ %\midrule
        \midrule%\midrule
		Field removed & 
		\(f \in \text{fields}(T) \land \nexists\, f' \in \text{fields}(T') \mid f \sim f'\) & \ding{55} & \ding{55} \\ %\midrule
		Field now protected & 
		\(f \sim f' \land \text{public}(f) \land \text{protected}(f')\) & \ding{55} & \ding{55} \\ %\midrule
		Field now final &
		\(f \sim f' \land \neg\text{final}(f) \land \text{final}(f')\) & \ding{55} & \ding{55} \\ %\midrule
		Field now static & 
		\(f \sim f' \land \neg\text{static}(f) \land \text{static}(f')\) & \ding{55} & \ding{51} \\ %\midrule
		Field no longer static & 
		\(f \sim f' \land \text{static}(f) \land \neg\text{static}(f')\) & \ding{55} & \ding{55} \\ %\midrule
		Field type changed & 
		\(f \sim f' \land \text{type}(f) \neq \text{type}(f')\) & \ding{55} & \ding{55} \\ \midrule
	\end{tabular}
\end{table*}

\subsection{Detecting Breaking Changes}
\label{sec:detection}
Detecting BCs between two library versions involves comparing their APIs $\mathcal{A}$ and $\mathcal{A'}$ to identify changes that disrupt backward compatibility, either at the source or binary levels.
%The JLS specifies minimum standards for binary compatibility~\cite[chap.~13]{jls} but does not explicitly list source-incompatible changes.

Our approach revolves around matching symbols pairwise across compared versions to analyze their differences.
We use the notation $s \sim s'$ to indicate that $s \in \mathcal{A}$ and $s' \in \mathcal{A'}$ share the same unique identifier in $\mathcal{A}$ and $\mathcal{A'}$.
If there is no symbol $s' \in \mathcal{A'}$ matching $s \in \mathcal{A}$ then $s$ has been removed from the API, which is always a BC.
If there is no symbol $s \in \mathcal{A}$ matching $s' \in \mathcal{A'}$ then $s'$ is a new symbol.
Inserting new symbols into an API is backward-compatible unless the new symbol is an abstract method, which requires clients to implement it.
Note that renaming a symbol, in terms of backward compatibility, constitutes a BC and is represented as both a removal (of the previous symbol) and an addition (of the new symbol).
Finally, when two symbols of the same kind match, their properties are checked for BCs using straightforward detection rules.
These detection rules are derived from the JLS~21~\cite{jls}, and we evaluate their accuracy in \Cref{sec:evaluation}.
Naturally, they are largely aligned with those found in \japi and \rapi, with the adjustments presented hereafter.
\Cref{tab:bcs} lists the BCs that can be detected atop API models, their associated detection rules, and whether they break binary or source compatibility.

A key difference between our approach and state-of-the-art tools such as \japi and \rapi is that BCs are expressed \emph{in terms of the API}, not in terms of the underlying source.
For instance, when the visibility of a field is decreased, \japi identifies the BC \textit{field less accessible} and \rapi the BC \textit{field visibility reduced}.
In contrast, because API models can only represent \ijava{public} and \ijava{protected} fields, the corresponding BC in our approach is called \emph{field now protected}.
When a field goes from \ijava{protected} to \ijava{private}, it is instead reported as a \emph{field removed} since the field is no longer part of the API, which better reflects how the API has evolved.
While both \japi and \rapi offer some support for detecting certain non-BCs (such as introducing a new \ijava{default} method in an interface), we focus strictly on BCs.

Although they are not explicitly referenced in detection rules, API-related constructs recently introduced in Java, such as \ijava{record} and \ijava{sealed}, are fully supported.
A \ijava{record} class is an implicitly final class that holds final fields known as record components~\cite[\S8.10]{jls}.
A \ijava{sealed} class or interface specifies all its direct subtypes when it is declared and cannot be extended beyond the closed world of the library.
These constructs are governed by the same rules as regular classes and interfaces with a few additional constraints, are affected by the same kinds of BCs, and require no further treatment.
We further evaluate the accuracy of our detection rules and their implementation in \Cref{sec:evaluation}.

\section{\rs: Fast and Accurate BC Detection}
\label{sec:roseau}

We have implemented our approach in a new static analysis tool, \rs.
To address the scenarios discussed in this paper, \rs is designed with three primary objectives:~high accuracy, excellent performance, and the ability to handle both bytecode and source code.

In a nutshell, \rs consists of three main components.
API extractors are responsible for turning library code, either in source code or bytecode form, into API models.
New extractors can be easily contributed by leveraging existing parsing technologies to build API models that conform to the rules expressed in \Cref{sec:approach}.
A diff component is responsible for comparing two API models and inferring a list of syntactic BCs following the rules listed in \Cref{tab:bcs}.
Finally, a report component formats the list of BCs into a machine-readable or user-readable report in JSON, CSV, or HTML.

\subsection{Backends}

\rs currently supports inferring API models from either source code or bytecode using two independent backends.
Each backend is responsible for parsing and visiting the source to extract the syntactic information necessary to build API models:
\begin{itemize}
	\item \emph{JDT}: The JDT parser is part of Eclipse's \textit{Java Development Tools} and is a popular solution for efficiently parsing Java source code.\footnote{\url{https://wiki.eclipse.org/JDT_Core/}} As it offers excellent performance, the JDT backend is best suited for large-scale analyses;
	\item \emph{ASM}: OW2's ASM is a bytecode analysis and manipulation framework focused on performance.\footnote{\url{https://asm.ow2.io/}} It is a widely-used framework, including as part of the OpenJDK and the Groovy and Kotlin compilers, and the preferred backend for bytecode analysis.
\end{itemize}

API models are self-contained and independent of the underlying source, so \rs automatically dumps the source ASTs or class files once the API model is constructed.
This significantly lowers the memory footprint of \rs, as this intermediate information can quickly fill up memory for large libraries~\cite{le2022hyperast}.
API models are lightweight and can be maintained in memory for further analyses while maintaining a minimal memory footprint (\eg, using them as a pivot when walking through a list of commits).

When parsing the source library, some references in API models might be dangling when \rs cannot resolve the corresponding type declaration.
This occurs when the API refers to types not provided in the classpath, \eg, when the API uses other third-party libraries.
To mitigate this issue, \rs natively resolves all symbols in Java's standard library and accepts a user-defined classpath as input when necessary to complete the resolution process.

APIs can be compared regardless of what they are extracted from:~an API inferred from a JAR release on Maven Central can be transparently checked for BCs against an API inferred from the source code of the latest commit fetched from GitHub.
APIs inferred from source code and from the equivalent compiled bytecode are identical, except for potentially missing information in bytecode depending on the compiler's configuration, which does not affect BC detection (\eg parameter names, precise position of the symbol in the source file, \etc).
Information about generic types remains accessible in bytecode despite erasure (using signatures~\cite[\S4.7.9]{jlsjvm}) and is incorporated in the API models by the ASM backend.
API models can also be serialized and deserialized as JSON when needed, enabling their reuse in subsequent analyses.
This is particularly handy for tagging a specific API version as a reference (\eg the latest release) and reusing it at virtually no cost.

Finally, to accurately infer the API of a software library, \rs implements the API delimitation mechanisms presented in \Cref{sec:background}.
Specifically, users can set up regex-based exclusions for packages (\eg \texttt{*.internal.*}), choose to ignore symbols marked with user-defined annotations (such as Google’s \texttt{@Beta} or Apache’s \texttt{@Internal}), and determine whether \texttt{@Deprecated} symbols should be included in backward compatibility checks.

Our implementation of API models offers several benefits over other tools.
For instance, \japi constructs an explicit \emph{change model} from two JAR files, serving as a thin layer of abstraction still closely tied to the underlying bytecode parser Javassist.
However, this model captures the differences between two APIs, not the API itself.
As such, it is specific to the comparison of two versions and does not allow reusing the API extracted from a given version for further analyses (\eg, against multiple other versions).
In contrast, our API model is fully self-contained and technology-independent, allowing an easier evolution with new versions of Java, the transparent definition of new backends, and the ability to maintain API models in memory or on disk for further analyses.

\subsection{Performance}
\label{sec:performance}

\rs is designed with a strong focus on performance, implementing several optimizations to enable large-scale studies of API evolution.

\paragraph*{Diet parsing}
In all backends, \rs prioritizes performance by parsing input source code or bytecode in ``diet'' mode.
This entails ignoring method bodies, expressions, and statements (such as field initializers) in source files, as these elements do not influence the API.
Similarly, code, debug, and frame attributes in bytecode are neither analyzed nor visited.
The JDT and ASM backends are configured to enable this ``diet'' mode.

\paragraph*{Type references and immutability}
API models use type references to link names to type declarations (classes, interfaces, type parameters, primitive and array types, \etc).
In \rs, these type references enjoy several properties.
Type references are strongly typed.
For instance, references to superclasses are type-guaranteed to refer to \ijava{class} symbols, not any other kind of type declaration.
They are lazily resolved only when needed for efficiency.
Additionally, they are unique within the typing scope of an API, meaning that there is only one single reference toward a given type declaration shared by all referencers.
A given name is thus only resolved once.
Finally, API models are implemented as fully immutable data structures, allowing full parallelization of the evaluation of the BC detection rules depicted in \Cref{tab:bcs}.
% \paragraph{Immutability}
% In \rs, API models are implemented as an immutable data structure, with the exception of type references that are resolved lazily.
% This enables 

\paragraph*{Incremental updates}
When comparing two versions of the same API, it is likely that they share most of their symbols.
This is typically the case when analyzing two subsequent releases or all commits within a given branch.
\rs is specifically designed to handle these scenarios by supporting incremental updates of API models.
This allows API models to be partially updated without parsing and inferring the API from scratch whenever the underlying source changes (\eg, a source tree or a JAR file).
API models can be updated incrementally to reflect only the changes between two library versions, drastically improving performance in scenarios where multiple versions share a large portion of their symbols.

Every symbol in the API model is linked to a physical location that identifies its source (a \texttt{.java} or \texttt{.class} file).
When the source is updated (\eg, when checking out a new commit), \rs computes the minimal set of files that have been added, deleted, or changed between the two versions.
This information can be extracted in various ways, for instance, by using hash-based comparisons, modification-time checks, or commit information from a version control system.

Once the changed files are identified, the update proceeds as follows:
\begin{itemize}
	\item \textit{Added files:} new files are parsed, and their corresponding symbols are integrated into the API model;
	\item \textit{Deleted files}: symbols originating from deleted files are removed from the API model. Assuming that the source is consistent (the source code compiles or the JAR file is valid), any reference to a deleted symbol must also have been updated in dependent files. Therefore, if a symbol is deleted, other files referencing it will have been modified and thus re-parsed to reflect the change;
	\item \textit{Changed files}: modified files are parsed again, and the corresponding symbols in the API model are updated accordingly. Since the API model uses name-based type references, any change to a symbol is automatically visible to all other symbols referencing it.
\end{itemize}

Note that since API models are immutable in \rs, unchanged symbols are first copied from the original API into the new version.
By leveraging incremental updates, our approach minimizes unnecessary re-parsing, ensuring that only symbols affected by changes in the source are recomputed.
As demonstrated in the next section, this drastically improves performance, especially when conducting large-scale longitudinal studies of library histories.

\section{Evaluation}
\label{sec:evaluation}

In our evaluation, we aim to assess the accuracy and runtime performance of \rs in detecting BCs, including in the context of large-scale longitudinal studies. 
Throughout the experiments, we compare \rs (version 0.1.0) against two baselines.
Originally, \citeauthor{jezek2017api} inventoried nine BC detection tools for Java libraries~\cite{jezek2017api}.
However, only two have been updated in the past five years and are still actively used in practice (\cf \Cref{sec:background}): \japi (version 0.23.1) and \rapi (version 0.15.0).
To the best of our knowledge, no other tool has emerged since \citeauthor{jezek2017api}'s inventory.

We evaluate each tool through three experiments.
The first evaluates their ability to accurately identify BCs (\Cref{sec:accuracy}).
The second measures their runtime performance on popular Java libraries (\Cref{sec:performances}).
The third uses Guava as a case study to evaluate their suitability for analyzing large commit histories (\Cref{sec:case-study}).

All experiments in this section were conducted on an Intel\textregistered{} Xeon\textregistered{} W-2125@4.00~GHz with 32~GB of RAM on Linux 6.1.0-generic using Java HotSpot 21.0.5 (2024-10-15 LTS).
The source code, scripts, and data discussed in this section are available in our replication package~\cite{latappy_replication_2025}.

\subsection{Accuracy}
\label{sec:accuracy}

\subsubsection*{Protocol}
To evaluate the accuracy of BC detection tools, we refine and extend the original benchmark of \citeauthor{jezek2017api}, which has already been used for accuracy evaluation in prior work (\eg, \cite{ochoa2022breaking}).
This benchmark follows a systematic methodology to list API evolution cases.
It considers possible changes along three dimensions: what is changing (\eg, a modifier), where the change is applied (\eg, a method), and how it is changed (\eg, it is removed), resulting in 224 possible combinations corresponding to changes (breaking and non-breaking) that may happen in real-world APIs~\cite{jezek2017api}.
Concretely, the benchmark comprises a first version of a synthetic API and a second version with the changes applied.
A third component, the client, contains a set of Java files.
Each file contains a \texttt{main()} method using a symbol that compiles successfully against the first API.
To build a ground truth, the benchmark automatically (i)~compiles all files in the client against the second API version to detect source-breaking changes and (ii)~links the existing client binary against the binary of the second API version to detect binary-breaking changes.
Finally, it runs BC detection tools on both versions of the APIs and compares their reports with the ground truth using a simple grep-based search on the client's file name to determine whether they accurately identify BCs.
The benchmark calculates the accuracy of each tool, defined as the ratio of correctly diagnosed changes over the total number of changes.

This benchmark is a solid foundation for evaluating BC detection tools.
However, it suffers from several drawbacks.
First, changes are sometimes intertwined, with the same file in the first version declaring several symbols all affected by different changes in the second version.
A single client sometimes uses several symbols:~if a linking error happens on the first, the linker aborts and subsequent errors are not evaluated for binary compatibility.
Following \citeauthor{ochoa2022breaking}'s approach, we refactor the benchmark so that only a single symbol is changed between the two versions of the same file, and each client only uses a single symbol of the API~\cite{ochoa2022breaking}.

Second, the use of shell scripts and \texttt{grep} to determine whether a tool's report matches the ground truth is brittle.
We find several instances where cases are mixed because they share a common prefix (\eg, \textit{genericsIfazeTypeAdd} and \textit{genericsIfazeTypeAddSecond}), which produces unsound results.
We address this problem by implementing a Java pipeline that programmatically runs each case independently, solving the unsoundness issues in the benchmark.

Third, the accuracy metric computed in the original benchmark confounds false positives and false negatives, and does not evaluate the tools' ability to distinguish source-breaking and binary-breaking changes, hindering the interpretability of the results.
In our refined benchmark, we evaluate the tools' precision and recall independently and report separate scores for their ability to distinguish source-breaking changes and binary-breaking changes.

Finally, we realize that the client code used as an oracle in the original benchmark is sometimes too weak. 
When a client fails to compile or link with the second version, the diagnosis is clear.
When a client successfully compiles and links, however, then either the change is indeed backward-compatible, or it is breaking but the client was unable to trigger the incompatibility.
For example, the \textit{genericsIfazeTypeBoundsAdd} case involves adding a bound to an interface's type variable in the API (\ijava{Lib<A>} to \ijava{Lib<A extends Number>}).
However, the corresponding client uses it with an \texttt{Integer} bound (\ijava{Lib<Integer>}), which does not trigger any compilation error when the client migrates to the second version, as \texttt{Integer} complies with the added bound.
This change is thus misclassified as a binary-compatible change.
A client using the interface with a \texttt{String} bound, for instance, triggers the incompatibility.
Thus, we manually review all clients in the benchmark to transform them into stronger oracles.
Notably, we addressed numerous cases related to type bounds and introduced new clients that extend types and override methods---in addition to simply instantiating and invoking them---to simulate inversion-of-control uses.
Our final benchmark contains 267 cases.
86 are backward-compatible, 81 are source-breaking, 21 are binary-breaking, and 79 are both source- and binary-breaking.

\begin{table}[!ht]
\setlength{\tabcolsep}{4pt}
\centering
\caption{Precision, recall, and \fs of \japi, \rapi, and \rs.}
\label{tab:accuracy}
\begin{tabular}{lrrrrrrrrr}
& \multicolumn{3}{c}{All} & \multicolumn{3}{c}{Source} & \multicolumn{3}{c}{Binary} \\
\cmidrule(lr){2-4}\cmidrule(lr){5-7}\cmidrule(lr){8-10}
& Prec. & Rec. & F\textsubscript{1} & Prec. & Rec. & F\textsubscript{1} & Prec. & Rec. & F\textsubscript{1} \\
\midrule
\japi & 0.90 & 0.82 & 0.86 & 0.78 & 0.80 & 0.79 & 0.90 & \textbf{0.98} & \textbf{0.94} \\
\rapi & 0.85 & 0.97 & 0.91 & 0.77 & 0.96 & 0.85 & 0.89 & 0.95 & 0.92 \\
\rs   & \textbf{0.98} & \textbf{0.99} & \textbf{0.99} & \textbf{0.88} & \textbf{1.00} & \textbf{0.94} & \textbf{0.91} & \textbf{0.98} & \textbf{0.94} \\
\end{tabular}
\end{table}

\subsubsection*{Results}
We use our benchmark to compute the accuracy of \rs, \japi, and \rapi.
As the results obtained on the benchmark using the two backends of \rs are identical, we do not distinguish them in this experiment.

\Cref{tab:accuracy} shows the precision, recall, and \fs obtained by the three tools.
Overall, all three obtain decent scores on the benchmark.
In all cases, \rs achieves slightly higher scores than the other tools, only tied in recall and \fs by \japi for binary-BCs.
While all tools are nearly as accurate on binary-BCs, the benefits of \rs are more apparent for source-BCs.
This might be explained by the original focus of \japi and \rapi on binary-BCs.
\japi, for instance, was originally focused on detecting the binary incompatibilities specified in the JLS~\cite[chap.~13]{jls} but has gradually evolved to support source incompatibilities.
Indeed, \japi achieves higher scores than in the original benchmark of \citeauthor{jezek2017api}:~at the time, it was not able to detect any BC related to generics;~this support was since implemented in the version used in our experiments.
As \rs was originally designed to analyze both source code and bytecode, we always considered both kinds of compatibility.
Overall, we find that \rs outperforms other tools in accuracy, with an overall \fs of 0.94 for source- and binary-breaking changes, and 0.99 for all changes---where the ability to differentiate source- and binary-breaking changes is not evaluated.

\begin{finding}
\textbf{Accuracy:}
\rs achieves higher overall accuracy ($F_1 = 0.99$) than \rapi ($F_1 = 0.91$) and \japi ($F_1 = 0.86$), primarily due to improved detection of source-breaking changes.
\end{finding}

\subsection{Performance}
\label{sec:performances}

In this experiment, we evaluate the runtime performance of the tools in a scenario where two versions of a library are compared for BCs.
This represents the typical use of BC detection tools as part of build managers or continuous integration.
As there is no past version of the library to consider, this experiment does not evaluate \rs's support for incremental updates, which we address in the last experiment.

The choice of backend has a significant impact on \rs's runtime performance.
The time spent building an API model and comparing two API models is negligible compared to the time spent parsing and visiting the source files or class files.
In this experiment, we thus consider two variants of \rs:~\rsb, which uses the ASM backend, and \rss, which uses the JDT backend.

\subsubsection*{Protocol}

To measure the performance of each tool in realistic conditions, we first curate a dataset of popular libraries.
First, we collect the top-100 artifacts with the most uses on Maven Central from \texttt{https://mvnrepository.com}.
We discard from this set all artifacts that are not libraries or do not contain any source code (\eg, Spring Boot starter projects) or that are implemented in other JVM languages (\eg, the standard libraries of Kotlin, Clojure, and Scala).
The resulting set contains 60 popular Java libraries, including h2, Spring, jackson-databind, assertj, commons-collections, \etc.
Finally, we include Java-21's standard library: 69 standard modules (\eg \texttt{java.lang}, \texttt{java.io}, \etc) totaling 4M lines of code.
We retrieve each library's source code and binary JAR from Maven Central, selecting only the most recent stable version.

We evaluate each tool's performance by comparing each library's last stable version with itself.
Although no BC should be detected in this scenario, it represents the worst-case scenario where the tools must traverse the entire API to thoroughly check for differences.
This prevents possible optimizations such as automatically marking all methods in a class as deleted when the class itself is deleted, without visiting them.
To ensure a fair comparison of the tools, we created a functionally equivalent Java wrapper for each.
This wrapper runs the analysis without generating a report, thereby avoiding any I/O beyond loading and analyzing the libraries.

We evaluate the runtime performance of \rsb, \japi, and \rapi (using the library's JAR as input and ignoring the potential compilation time necessary to build it), as well as of \rss (using the library's sources as input).
This experiment does not exercise \rss's incremental updates, which we evaluate in the next experiment.

For accurate and stable measurements, we use a JMH harness\footnote{\url{https://openjdk.org/projects/code-tools/jmh/}} configured with 5 fresh JVM forks and 10 measurements in single-shot time mode to obtain 50 measurements per library and tool.
We set the warm-up iterations to 10 as we empirically verified that all tools reach stable performance at that point.

\subsubsection*{Results}

\begin{figure}
    \centering
    \includegraphics[width=\linewidth]{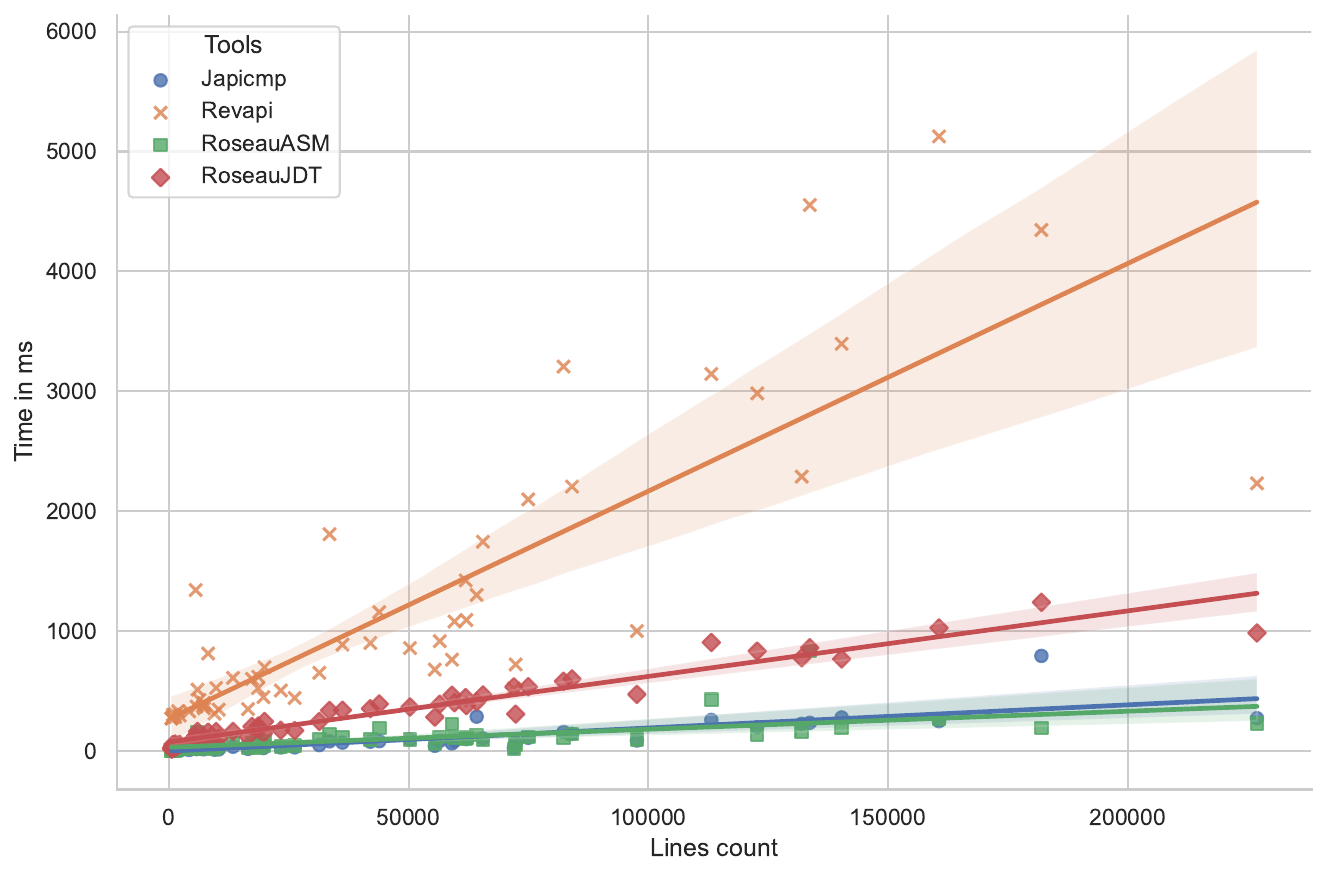}
    \caption{Execution time of BC detection tools across 60 popular Java libraries.}
    \label{fig:lines_impact_warm}
\end{figure}

\Cref{fig:lines_impact_warm} shows the relation between the tools' runtime performance and the size of the analyzed libraries (lines of code as reported by the \texttt{cloc} utility).
In all cases, there is a clear linear relation between library size and the tools' runtime performance, although \rapi's data points are more scattered.
When analyzing JARs, \japi and \rsb perform similarly, inferring BCs for all libraries up to 250k lines of code in less than a second.
\rapi, however, is visibly slower than the other tools.

\rss is the only tool able to analyze source code directly.
Surprisingly, it can analyze all 60 libraries in less than 2~seconds, mainly thanks to the optimized ``diet'' parsing implemented in \rs.
In scenarios where only source code is available, \rss largely surpasses the other tools as they would require compiling the source code first.
We evaluate and confirm this intuition in the next experiment.

Regarding the standard library of Java~21 and its 4M lines of code (not included in \Cref{fig:lines_impact_warm} for the sake of readability), the mean analysis time is 5,157~ms for \japi, 148,535~ms for \rapi, 4,834~ms for \rsb, and 18,990~ms for \rss.
Here again, \japi and \rsb perform similarly when analyzing the JAR version of the JDK, with \rapi lagging behind.
\rss is able to parse the 4M lines of code, build the API model, and conduct the analysis in 19~seconds, showcasing its potential for very large libraries.

This demonstrates that \rs can analyze libraries with excellent performance regardless of their size, both as JARs or in their source code form.
\rs can thus serve as a drop-in replacement for \japi and \rapi in continuous integration and build managers with better accuracy, and opens new analysis capabilities that we evaluate in the next experiment.

\begin{finding}
\textbf{Performance:}
When analyzing JARs, \rs performs similarly to \japi in terms of execution times, analyzing libraries up to 250k lines of code in less than a second.
When analyzing source code, \rs largely outperforms \japi and \rapi (which require compilation), analyzing all libraries in less than 2 seconds.
\rs parses and analyzes the source code of Java 21's entire standard library (4M lines) in under 19 seconds.
\end{finding}

\subsection{Case Study}
\label{sec:case-study}

\begin{figure*}[ht]
    \centering
    \includegraphics[width=\linewidth]{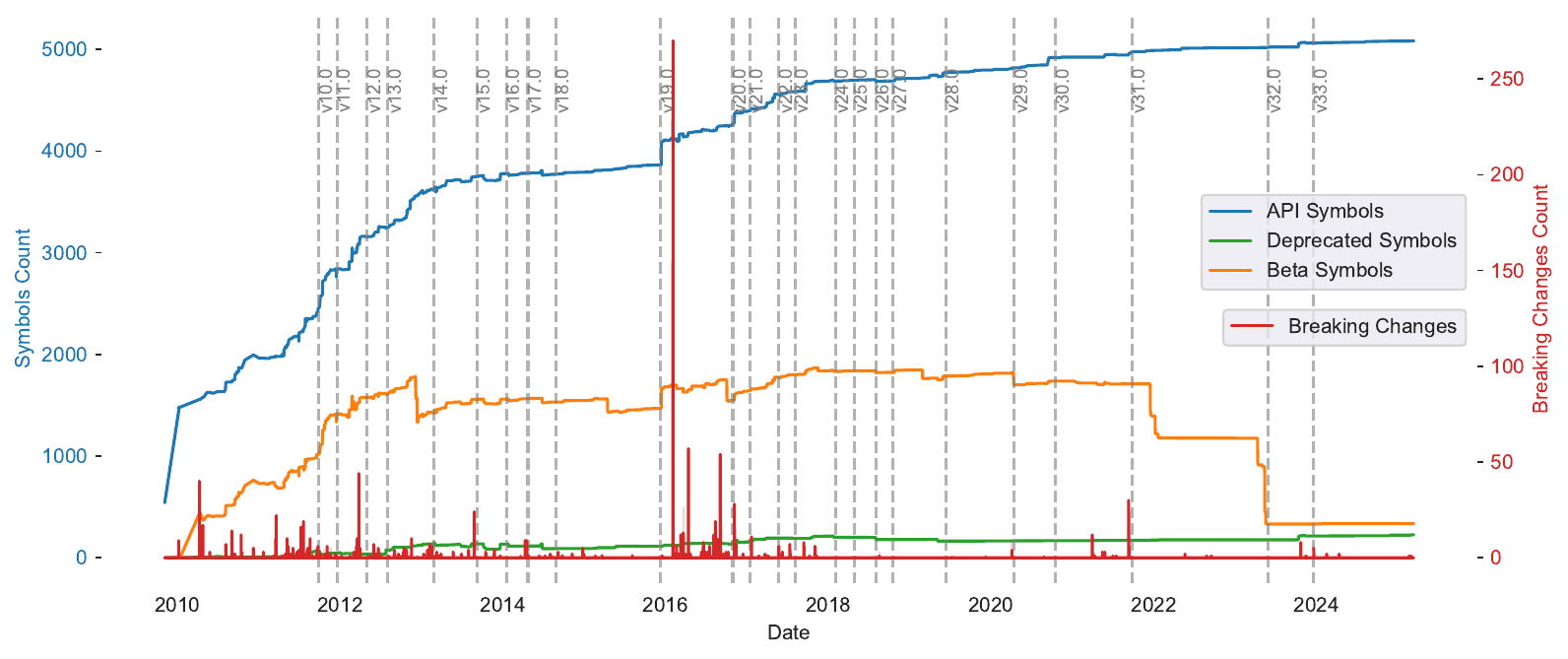}
    \caption{Evolution of Guava's API and introduction of BCs over time. Major versions are depicted as vertical dashed lines.}
    \label{fig:guava-api}
\end{figure*}

In this case study, we aim to evaluate the tools' suitability to conduct large-scale longitudinal studies using Google's Guava library as the subject.
Guava is a very popular library with a large number of contributors and, at the time of writing, a rich history of 6,839 commits spanning 14 years.
In this case study, our goal is to analyze sequentially all commits in Guava's main branch throughout its evolution.
We limit our analysis to Guava for brevity, and refer the reader to our replication package for a similar analysis of \texttt{commons-lang} and \texttt{h2database}.

\subsubsection*{Protocol}

To illustrate the suitability of each tool for this task, we first conduct a preliminary assessment restricted to the latest 100 commits on Guava's main branch.
Our objective is to illustrate the amount of time required by \rs, \japi, and \rapi to analyze these 100 commits.
Retrieving them is straightforward, but they only exist in the form of source code.
As \japi and \rapi require binary JARs as input, we first measure the time necessary to compile and package these 100 commits as JAR files without running any tests.
Then, we measure the time taken by \rss, with and without incremental updates, to build the API model of the commit.
In both cases, when comparing commits $c_{i-1}$ and $c_i$, the API model of $c_i$ is kept in memory to be reused as a pivot when comparing $c_i$ and $c_{i+1}$.
To estimate the time each tool would have taken to analyze Guava's entire commit history, we calculate the average time required to analyze each of the 100 commits and multiply it by the total number of commits.

In a second step, we use the most efficient approach, \rs with incremental updates, to analyze all commits in Guava's history.
We present the extracted information, focusing on the evolution of its API and the introduction of BCs over time.
We detail some key events in its history, showcasing the suitability of \rs for conducting empirical studies of library histories.

\begin{table}[ht]
\setlength{\tabcolsep}{3pt}
\centering
\caption{Mean time and estimated total time to build a JAR or extract an API model from a sample of 100 Guava commits.}
\label{tab:preliminary}
\begin{tabular}{lrr}
Method & Mean commit time (ms) & Total time (minutes) \\
\midrule
Packaging JAR & $68,750$ & $7,836$ \\
Non-incremental \rs & $1,065$ & $121$ \\
Incremental \rs & $81$ & $9$ \\
\end{tabular}
\end{table}

\subsubsection*{Results}

\Cref{tab:preliminary} depicts the results of our preliminary experiment.
On average, packaging Guava's source code to obtain a JAR takes 69~seconds per commit.
Analyzing the history of Guava would thus take nearly five and a half days using \japi or \rapi.
This highlights their limitations for conducting large-scale longitudinal studies.
Without incremental updates, \rs takes about a second to build an API model from each commit and roughly two hours to analyze all commits.
Finally, \rs performs best when incremental updates are activated, as it minimizes the amount of source code to parse at each step.
In this configuration, \rs takes 81~ms per commit, or 9~minutes for all commits.

Finally, we end the preliminary experiment and use \rs with incremental updates to analyze all 6,839 commits in Guava's history and compute the BCs between each pair of subsequent commits.
On average, \rs analyzes a new commit in 117~ms---this varies with the number of added, changed, and removed files in each commit due to incremental updates.
This is slower than in the preliminary experiment, as this includes the time taken to identify BCs and compute metrics.
\rs analyzes all 6,839 commits in Guava in 799~seconds (approximately 13~minutes): 208~seconds walking from one commit to the next using \texttt{git checkout} (26\%); 284~seconds parsing source code and inferring the API model (36\%); 247~seconds diffing two APIs to identify BCs (31\%); and 60~seconds computing metrics and storing the results.
One key benefit of \rs is that intermediate API models can be reused in the next step:~for each commit, it compares the new version with the previous one stored in memory.
This demonstrates that, unlike existing tools, \rs enables large-scale longitudinal studies of API evolution and BCs.

\Cref{fig:guava-api} shows the information \rs extracts at each analysis step and highlights key moments in Guava's API evolution.
Commits are represented on the X-axis. For each commit, the plot shows the number of API symbols (including types, executables, and fields), the number of \texttt{@Beta} and \texttt{@Deprecated} symbols, and the amount of BCs the commit introduces.
This analysis is possible thanks to \rs's reified API models that can not only detect BCs, but also track API evolution, in contrast with \japi and \rapi.
Without claiming to conduct an in-depth qualitative analysis, we discuss some interesting moments in Guava's history to showcase \rs's value for empirical research on API evolution.

A first observation is that Guava's API has steadily grown over time, doubling the number of exported API symbols between versions 10.0 and 33.0.
From the start, a significant portion of Guava's API was exposed as \texttt{@Beta}, following the maintainers' guidelines.
In version 32.0, however, Guava's maintainers decided to freeze most of their beta APIs: ``\textit{the `@Beta` annotation is removed from almost every class and member. This makes them officially API-frozen.}'',\footnote{\url{https://github.com/google/guava/releases/tag/v32.0.0}} as clearly shown in \Cref{fig:guava-api}.
Regarding BCs, most were introduced in earlier versions of the library, with very few BCs introduced after 2018, highlighting Guava's maturity.
On Feb 2, 2016, shortly after the release of version 19.0, commit \texttt{2067e7} introduced 270 BCs.
This commit rolled back a set of overloads introduced in \texttt{com.google.common.base.Preconditions} that were introduced to avoid autoboxing and numeric conversions.
The rollback commit thus removed 270 methods from this class.
Version 20.0 and 31.0 also included a notable number of BCs, the release notes indeed discussing removed APIs.

\begin{finding}
\textbf{Large-scale longitudinal studies:}
\rs analyzes all 6,839~commits of the Guava library, spanning 14~years, in 13~minutes.
Its reified API models enable commit-level tracking of API evolution and breaking changes, making it ideal for large-scale empirical studies of API evolution.
\end{finding}

%At first, we had the idea to create a new benchmark with a combinatorial approach. Based on previous studies on Java API~\cite{monce_lightweight_2024, qiu_understanding_2016}, we have the list of all valid interactions for each kind of symbol. Using some brut-force, we could generate all permitted interactions from the JLS~\footnote{Example for Java 21: \url{https://docs.oracle.com/javase/specs/jls/se21/html/index.html}} to build one first API version, then use those listed interactions to generate all clients consuming this API and finally, applying all possible atomic code changes to the first API version and obtain a set of new versions. It simulates a real scenario where one client uses a given version of an API and this version is updated. The Java Compiler permits to build the ground truth and we could evaluate our selected tools by comparing the two generated versions. Yet, already at a reduced generation scale the combinatorial explodes. For one generated API with around 300 types, 45K methods and 2K fields, we were building almost 48K clients and we estimated at 1.5M the number of code changes possible for this API. For now, this approach is not feasible but represents an interesting future work.

\subsection{Threats to Validity}

\subsubsection*{Construct Validity}
In \Cref{sec:accuracy}, we refine the clients to strengthen their ability to detect BCs in the two API versions.
We manually review all changes and corresponding clients in the benchmark, but we cannot guarantee that no change was mislabeled.
While the original benchmark was constructed systematically, exhibiting most of the possible changes to Java APIs, we cannot guarantee that it is comprehensive, especially when considering multiple interacting changes occurring concurrently.
Therefore, it is possible that some changes that could happen in real-world libraries are not covered in the benchmark.

\subsubsection*{Internal Validity}
Accurate runtime measurements are challenging and can be affected by many external factors~\cite{costa2019s}.
We used a dedicated benchmark framework, JMH, and a dedicated machine with only the required dependencies that was not performing any other task during the experiments.
In \Cref{sec:case-study}, we assume that the time taken to compile and package the last 100 commits of Guava can be used to extrapolate compilation times throughout its history.
While this is an over-approximation (earlier versions were smaller and likely to compile faster), this does not jeopardize our conclusions. 

\subsubsection*{External Validity}
In \Cref{sec:performances}, we curate a set of 60 popular libraries for performance measurements.
While this set contains libraries built by diverse contributors and focusing on diverse domains, they are not a random sample of Java libraries.
The results may vary in another set of libraries.
Similarly, our case study uses Guava as the sole subject and may not be representative of the results obtained for other libraries.

\section{Conclusion}
\label{sec:conclusion}

In this paper, we have presented \rs, a new static analysis tool for Java libraries that infers accurate API models from either source code or bytecode to support API evolution analysis.
API models of two library versions can be compared to identify syntactic breaking changes between them.
Using an established benchmark of breaking changes, we have shown that \rs is more accurate than the state-of-the-art tools \japi and \rapi.
We have evaluated the performance of \rs and shown that it matches their performance when analyzing bytecode, and outperforms them by two orders of magnitude when analyzing source code, showcasing the suitability of \rs for large-scale longitudinal studies of API evolution, taking minutes instead of days.

As future work, we will incorporate \rs as part of IDEs to connect it to their internal code representation and provide live feedback on API design and evolution to library designers.
We will leverage the capabilities of \rs to conduct large-scale empirical studies of popular libraries to better understand the dynamics of client-library co-evolution and inform code reviews by providing maintainers with detailed API reports.

\section*{Data Availability}
\rs is hosted on GitHub (\url{https://github.com/alien-tools/roseau}).
This paper's experiments use version 0.1.0 (\url{https://github.com/alien-tools/roseau/releases/tag/v0.1.0}).
The replication package on Zenodo~\cite{latappy_replication_2025} contains the extended version of \citeauthor{jezek2017api}'s benchmark used in the accuracy evaluation, the 60 popular Java libraries used in the performance evaluation, the full experimentation pipeline, the raw results presented in \Cref{sec:evaluation}, and the associated plots and analyses.

\section*{Acknowledgments}
This work was partially funded by the French National Research Agency through grant ANR ALIEN (ANR--21--CE25--0007).
The authors thank Yasmine Hamdaoui for her initial work on \rs and the anonymous referees for their precious comments.

\balance
\bibliographystyle{IEEEtranN}
\bibliography{main}

% \cleardoublepage
% \appendix
% \section*{Follow-ups}

% \subsection{A Year of Library Evolution}
% \begin{itemize}
%     \item Doesn't have to be a year ;)
%     \item Case study: a few popular software libraries
%     \item Analyze every commit, issue, pull request, branch
%     \item How do libraries evolve over time? API size, breaking, refactorings, versioning, \etc
%     \item Who contributes to the API? Maintainers vs external contributions
%     \item What's missing for them?
% \end{itemize}

% \subsection{Live API Design}
% \begin{itemize}
%     \item El famoso IntelliJ plug-in
% \end{itemize}

\end{document}